# Social Protocols for Agile Virtual Teams

Willy Picard

Dept. of Information Technology, Poznan University of Economics,
Mansfelda 4, 60-854 Poznan, Poland
Willy.Picard@ue.poznan.pl

**Abstract.** Despite many works on collaborative networked organizations (CNOs), CSCW, groupware, workflow systems and social networks, computer support for virtual teams is still insufficient, especially support for agility, i.e. the capability of virtual team members to rapidly and cost efficiently adapt the way they interact to changes. In this paper, requirements for computer support for agile virtual teams are presented. Next, an extension of the concept of social protocol is proposed as a novel model supporting agile interactions within virtual teams. The extended concept of social protocol consists of an extended social network and a workflow model.

## 1 Introduction

Computer support for Human-to-Human (H2H) interactions has a long history in computer science: from early visionary ideas of Douglas Engelbart at the Stanford Research Institute's Augmentation Research Center on groupware in the 60's, through CSCW and workflows in the 80's, and with social network sites in the 2000's. However, computer support for agile H2H interactions is still insufficient in most collaborative situations.

Among various reasons for the weak support for H2H interactions, two reasons may be distinguished: first, many *social elements* are involved in the H2H interaction. An example of such a social element may be the roles played by humans during their interactions. Social elements are usually difficult to model, e.g. integrating hierarchical relations among collaborators to collaboration models. A second reason is the *adaptation* capabilities of humans which are not only far more advanced than adaptation capabilities of software entities, but also are not taken into account in existing models for collaboration processes.

The insufficient support for human-to-human interactions over a network is a strong limitation for a wide adoption of *professional virtual communities* (PVCs). As mentioned in [1], "professional virtual community represents the combination of concepts of virtual community and professional community. Virtual communities are defined as social systems of networks of individuals, who use computer technologies to mediate their relationships. Professional communities provide environments for professionals to share the body of knowledge of their professions […]". According to Chituc and Azevedo [2], little attention has been paid to the social perspective on Collaborative Networks (CN) business environment, including obviously professional virtual communities in which social aspects are of high importance. Additionally, the adaptation capabilities of humans have been the object of few works [3]. As a consequence, support for *agile virtual teams* (VT) is currently insufficient.

Virtual team agility (VTA) refers to the capabilities of a group of human beings, the VT members, to rapidly and cost efficiently adapt the way they interact to changes. Changes may occur:

- within the VT: e.g., a collaborator may be temporary unavailable or he/she may acquire new skills,
- in the environment of the VT: e.g., a breakdown of a machine may occur, weather conditions may prevent the realization of a given task.

In this paper, we present a model which provides support for agile VTs based on the concept of social protocols. In Section 2, requirements for a computer support for agile VTs are presented. Next, the concept of social protocols supporting agile VTs is detailed. The proposed solution is then discussed. Section 5 concludes the paper.

## 2   Requirements for Support for Agile Virtual Teams

### 2.1   A Model of the Social Environment

A first requirement for support for agile VTs is the modeling of the *social environment* within which interactions take place. Each VT consists of at least two members, each of them having her/his own social position. By social position, we mean a set of interdependencies with entities (generally individuals or organizations): e.g. a VT member has colleagues, works in a given company, and belongs to a family.

VTA implies a rapid adaptation of the VT to new conditions. The social environment is a core tool in the adaptation process as it provides information about available resources VT members are aware of:

- within the VT: e.g., if a VT member is temporary unavailable, another person in the social environment may substitute for the unavailable VT member,
- in the environment of the VT: e.g., if weather conditions prevent the realization of a given task, new VT members which were not initially involved in the realization of the cancelled task may be needed to overcome it.

A partial answer to the question of modeling a social environment may be found in popular in the last five years social network sites, such as LinkedIn [4], MySpace [5], Orkut [6], Facebook [7], to name a few. Boyd and Ellison [8] define social network sites as "web-based services that allow individuals to (1) construct a public or semi-public profile within a bounded system, (2) articulate a list of other users with whom they share a connection, and (3) view and traverse their list of connections and those made by others within the system." The second and third points of this definition illustrate a key feature of social network sites, i.e. social network sites allow users for an easy access to information about persons they know (friends, colleagues, family members) and potentially about contacts of these persons.

However, the model of social environment adopted in social network sites captures only interdependencies among individuals or organizations. The interdependencies with information systems, e.g. web services, are an important element of the landscape of interactions within VTs: while individuals represent the "who" part of the interactions, information systems usually represent the "how" part. A VT member (the individual) performs some activity with the help of a tool (the information system). Therefore, we claim that a model of the social environment for interactions within VTs should integrate both interdependencies among VT members and interdependencies among VT members and information systems.

Such a model of social environment would allow VT members to react to new situations not only by changing the set of members but also by changing the set of tools. Additionally, such a model would allow VT members for agility with respect to changes related with information systems: e.g., if an information system is unavailable, VT members may seek for an alternative in their social environment.

It should be noticed that, while the social environment encompasses the professional virtual community (PVC), some elements of the social environment can be external to the PVC. During the adaptation process of VTs, the identification of required resources, either VT members or information systems, should not be limited to the PVC, as some valuable resource may come from personal relations of VT members, external to the PVC.

### 2.2 Structured Interactions within Virtual Teams

Supporting agile VTs requires guidance for VT members about tasks they may perform at a given moment of time. Such a guidance allows VT members for *focusing on appropriate tasks* that need to be fulfilled at a given moment of time, in a given collaboration situation, instead of facing all potential tasks that they may perform.

The tasks that a given VT member may perform depend also on the *role* he/she is playing within a given VT. Therefore support for VTA implies the mapping between VT members and roles they are playing within a given VT. Additionally, interactions within VTs are often structured according to collaborative patterns [9, 10]. In similar situations, in different VTs, members perform activities whom successiveness is identical among the various VTs: e.g., a brainstorming session consists of 5 phases:
  1. the chairman presents the problem,
  2. every participant presents his/her ideas,
  3. the chairman classifies the ideas,
  4. every participant may comment any idea,
  5. the chairman summarizes the brainstorming session.

In the former example, each phase may be decomposed as a sequence of activities to be performed, with activities associated to roles. Interactions within VTs could therefore be structured with the help of a *process* and an *associated process model* specifying the sequences of activities, the association between activities and roles, and the mapping between VT members and roles.

Results of studies in workflow technology and process modeling [11 – 14] provide a strong foundation for support for structured interactions within VTs based on the concepts of workflow and process models.

### 2.3 Layered Interaction Models

The concept of process model presented in the former subsection as a mean to structure interaction within VTs has to be considered at three levels of abstraction:

- *abstract process model*: a process model is abstract if it defines the sequence of activities to be potentially performed by VT members playing a given role, without specifying neither the implementation of activities, nor the attribution of roles to VT members. As an example, an abstract process model for a brainstorming session may specify that, first, a chairman presents the brainstorming session problem, next, participants present their ideas. Neither the implementation of the presentation of the problem and participants' ideas, nor the VT members are defined in the abstract process model.
- *implemented process model*: a process model is implemented if it defines the implementation of activities defined in an associated abstract process model. As an example, an implemented process model based on the brainstorming abstract process model formerly presented may specify that the presentation of the brainstorming session problem will be implemented as an email to all par-ticipants, while the presentation of ideas will be performed as posts to a forum.
- *instantiated process model*: a process model is instantiated if the attribution of roles to VT members for a given implemented process model has been set. Additionally, an instantiated process model, referred also as *process instance*, keeps trace of the current state of the interactions within a given VT. As an example, *the* former implemented process model may be instantiated by specifying who plays the chairman role and who the participants are. Additionally, the process retains its current state which may for instance be "participants are presenting ideas".

The following analogy with object-oriented programming illustrates the three levels of abstraction presented above:

- abstract process models are similar to interfaces or abstract classes. An abstract process model does not rely, nor provide an implementation of activities, as an interface does not provide an implementation of methods;

- implemented process models are similar to classes. An implemented process model provides an implementation of activities, as a class provides an implementation of methods.
- instantiated process models are similar to objects. An instantiated process model rules the interactions according to a given implemented process model and has its own state, as an object behaves according to its class and has its own state.

The separation of these three levels of abstraction leads to *process model reuse*. By separating the logical structure of interactions from its implementation, an abstract process model may be reuse in various contexts, IT environments, VTs. The PVC may provide its members access to a library of abstract and implemented process models. As a consequence, VT members facing some unpredicted situation may identify an already defined abstract or implemented process model allowing them to solve their problem. Then, the VT may react rapidly by just (eventually implementing and) instantiating the process. The brainstorming process presented above is an example of an abstract or implemented process that may be reuse by various VTs in a given PVC to interact in an agile way.

### 2.4 Adaptability

Adaptability is a core requirement of support for VTA. Adaptability refers in this paper to the capability of a VT to modify *at run-time* the model ruling its interactions.

In typical workflow management systems, two parts may be distinguished: a *design time* part allows for definition of workflow schemas while the *run-time* part is responsible for execution of workflow instances. A main limitation of typical workflow management systems is the fact that once a workflow schema has been instantiated, the execution of the workflow instance must stick to the workflow schema till the end of the workflow instance execution.

PVCs are a typical case of environments in which there is a strong need for the possibility to modify a workflow instance at run-time. Such modifications are usually needed to deal with situations which have not been foreseen nor modeled in the associated workflow schema. Adaptability refers to the possibility to modify a running instantiated process model to new situations which have not been foreseen and modeled in the associated abstract/implemented process model.

## 3 Social Protocols

Computer support for VTA requires novel models to support requirements presented in Section 2. The solution presented in this paper is based on the concept of *social protocol*. This concept has been presented first in 2006 [15], based on the concept of *collaboration protocol* [3]. A generic extended version of the concept of social protocol, including elements related with the modeling of the social environment, has been formally presented in [16]. The application of extended social protocols to PVCs and VTs is presented in this section.

### 3.1 Abstract Social Protocols

An abstract social protocol, $SP_a$, consists of two parts:
- an *abstract social network*: a direct graph modeling interdependencies among *abstract resources*. An abstract social network models the social environment required for a particular collaboration pattern.
- an *abstract interaction protocol:* a direct graph modeling interdependencies among *abstract activities*. An abstract interaction protocol models the sequence of activities in a particular collaboration pattern.

An example of an abstract social protocol for brainstorming is presented in Fig. 1.

In an abstract social network, vertices represent abstract resources that may support or be actively involved in the collaboration process, such as a collaboration role or a class of information systems. Edges represents relations between resources associated with social interaction types, such as "works with", "has already collaborated with" among roles, or "is the owner", "uses" between a role and a class of information systems. Labels associated with edges are not predefined, as the concept of social protocol should be flexible enough to encompass new types of interdependencies among resources. Therefore, new labels may be freely created at design time.

In an abstract interaction protocol, vertices represent:
- abstract activities that may be performed during the collaboration process, such as "present the brainstorming problem" or "present an idea". Activities are associated with a given role, e.g. only the chairman may present the brainstorming problem;
- states in which the group may be at various moments of the collaboration process, e.g. the group may be "waiting for ideas".

Edges run between activities and states, never between activities nor between states. Edges capture the potential activities in a given state, or states after the execution of a given activity. One may recognize in abstract interaction protocols the concept of Petri nets, where states are places and activities/roles pairs are transitions.

### 3.2 Implemented Social Protocols

Similarly to the relation between implemented process models and abstract process models presented in Section 2.3, an implemented social protocol defines the implementation of abstract activities associated with an abstract social protocol.

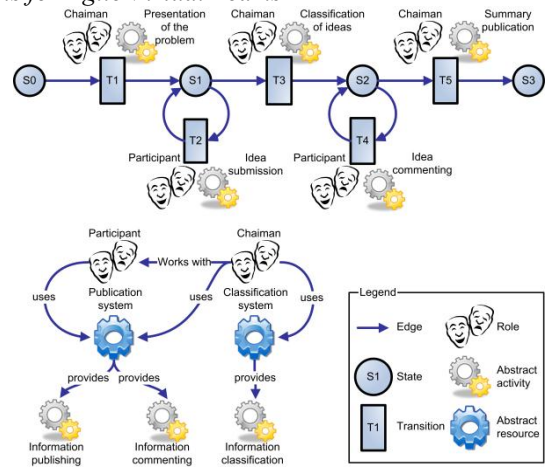

**Fig. 1.** An example of an abstract social protocol. At the top, the abstract interaction protocol of a brainstorming session. At the bottom, the abstract social network.

Therefore, an implemented social protocol consists of three parts:
- an abstract social protocol,
- a mapping of *abstract resources* associated to with abstract activities to *implemented resources*. For instance, the abstract resource "Publication system" of the former example may be mapped to a forum system on a given server.
- a mapping of *abstract activities* to *implemented activities*. For instance, the abstract activity "presentation of the problem" of the former example may be mapped to the URL of the form used to post information on the formerly mentioned forum system.

These two mappings may be built based on a pre-existing social environment defining interdependencies among resources (abstract and implemented). Additionally, the pre-existing social environment may be extended by the addition of missing resources. Therefore, on the one hand, the implementation procedure may take advantage of the social environment, on the other hand, the social network may benefit from the implementation procedure.

### 3.3 Social Processes

Similarly to the relation between instantiated process models and implemented process models presented in Section 2.3, a social process defines the implementation of abstract roles associated with an implemented social protocol, as well as keeps trace of the state of the interactions within the VT.

Therefore, a social process consists of three parts:
- an implemented social protocol,
- a mapping of *abstract resources* associated with roles to *collaborators*. For instance, the abstract resource "brainstorming chairman" is mapped to "John".
- a *marking* of active states.

The role-collaborator mapping may be built based on the pre-existing social environment. Additionally, the pre-existing social environment may be extended by the addition of missing resources, by the addition of collaborators. Therefore, on the one hand, the instantiation procedure may take advantage of the social environment, on the other hand, the social network may benefit from the instantiation procedure.

### 3.4 Meta-Processes

The concept of *meta-process* is our answer to the adaptation requirement. During the execution of an instantiated social protocol, collaborators may identify a need for modification of the process instance they are involved in. As a consequence, collaborators need to interact to decide how the process should be changed. A meta-process is a social process associated with another social process $\pi$ allowing collaborators of $\pi$ to decide in a structured collaborative way how the process $\pi$ should be modified. More information about meta-processes and adaption may be found in [16, 17].

## 4 Discussion

Some interesting works have been done in the field of electronic negotiations to model electronic negotiations with the help of negotiation protocols. In [18], it is stated in that, in the field of electronic negotiations, "the protocol is a formal model, often represented by a set of rules, which govern software processing, decision-making and communication tasks, and imposes restrictions on activities through the specification of permissible inputs and actions". One may notice the similarity with the concept of social protocol. The reason for this fact is that the model presented in this paper was originally coming from a work on protocols for electronic negotiations [15]. However, to our knowledge, none of the works concerning negotiation protocols provides support for the modeling of the social environment. Moreover, these works are by nature limited to the field of electronic negotiations which is just a subset of the field of interactions within VT.

As process modeling is concerned, many works have already been conducted in the research field of workflow modeling and workflow management systems. Many works [19 – 22] have focused on formal models and conditions under which a modification of an existing – and potentially running – workflow retains workflow validity, the ADEPT2 project [24] being probably the most advanced one. However, current works concerning workflow adaptation focus on interactions, and the importance of social aspects are not or insufficiently taken into account by these works.

Sadiq and al. [25] have proposed an interesting model for flexible workflows, where flexibility refers to "the ability of the workflow process to execute on the basis of a loosely, or partially specified model, where the full specification of the model is made at runtime, and may be unique to each instance." However, support for flexibility does not ensure support for adaptability, as flexibility, as proposed by Sadiq and al., implies that the workflow designer has specified at design time frames and boundaries to possible modifications of the workflow.

## 5 Conclusions

While many works are currently done on modeling collaboration processes in which software entities (agents, web services) are involved, modeling collaboration processes in which mainly humans are involved is an area that still requires much attention from the research community. Some of the main issues to be addressed are the social aspects of collaboration and the adaptation capabilities of humans. In this paper, the requirements of computer support for virtual team agility (VTA) are presented. Additionally, the concept of social protocol, combining social networks and workflow models, is proposed as a model supporting interactions within agile VT.

The main innovations presented in this paper are 1) the requirements for VTA, 2) the refinement of the concept of social protocol by the addition of a social network as a way to model the social environment, and 3) the three-layer view on social protocols – abstract, implemented, and instantiated – and the concept of meta-process.

A prototype is currently under implementation to validate the model presented in this paper. Among future works, methods to update the social network to reflect interactions within the VT performed in a given process are still to be proposed.

**Acknowledgments.** This work has been partially supported by the Polish Ministry of Science and Higher Education within the European Regional Development Fund, Grant No. POIG.01.03.01-00-008/08.

## References

1. Camarinha-Matos, L.M., Afsarmanesh, H., Ollus, M.: ECOLEAD: A Holistic Approach to Creation and Management of Dynamic Virtual Organizations. In: 6[th] IFIP Working Conf. on Virtual Enterprises, pp. 3--16, Springer (2005)

2. Chituc, C.M., Azevedo, A.L.: Multi-Perspective Challenges on Collaborative Networks Business Environments. In: 6$^{th}$ IFIP Working Conf. on Virtual Enterprises, pp. 25--32, Springer (2005)
3. Picard, W.: Modeling Structured Non-monolithic Collaboration Processes. In: 6$^{th}$ IFIP Working Conf. on Virtual Enterprises, pp. 379--386, Springer (2005)
4. LinkedIn, http://www.linkedin.com/
5. MySpace, http://www.myspace.com/
6. Orkut, http://www.orkut.com/
7. Facebook, http://www.facebook.com/
8. Boyd, D.M., Ellison, N.B.: Social Network Sites: Definition, History, and Scholarship. J. of Computer-Mediated Communication. 13(1), 210--230 (2007)
9. van der Aalst, W.M.P., van Hee, K.M., van der Toorn, R.A.: Component-Based Software Architectures: A Framework Based on Inheritance of Behavior. BETA Working Paper Series, WP 45, Eindhoven University of Technology, Eindhoven (2000)
10. Russell, N., ter Hofstede, A.H.M., Edmond, D., van der Aalst, W.M.P.: Workflow Resource Patterns. BETA Working Paper Series, WP 127, Eindhoven Univ. of Technology (2004)
11. Fisher, L.: BPM & Workflow Handbook. Future Strategies Inc. (2007)
12. Jeston, J., Nelis, J.: Business Process Management, Second Edition: Practical Guidelines to Successful Implementations. Butterworth-Heinemann (2008)
13. Harrison-Broninski, K.: Human Interactions. Meghan-Kiffer Press (2005)
14. van der Aalst, W. M. P., van Hee, K.: Workflow Management: Models, Methods, and Systems (Cooperative Information Systems). The MIT Press (2004)
15. Picard, W.: Computer Support for Adaptive Human Collaboration with Negotiable Social Protocols. In: 9$^{th}$ Int. Conf. on Business Information Systems, Lecture Notes in Informatics, vol. P-85, Gesellschaft fur Informatic, pp. 90--101 (2006)
16. Picard, W.: Computer Support for Agile Human-to-Human Interactions with Social Protocols. In: 12$^{th}$ Int. Conf. on Business Information Systems, LNBIP 21, pp. 121--132 (2009)
17. Picard, W.: Support for adaptive collaboration in Professional Virtual Communities based on negotiations of social protocols. Int. J. Inf.Tech and Mgmt, 8(3), 283--297 (2009)
18. Kersten, G.E., Strecker, S.E., Lawi, K.P.: Protocols for Electronic Negotiation Systems: Theoretical Foundations and Design Issue. In: 5$^{th}$ Conf. on Electronic Commerce and Web Technologies, pp. 106--115, IEEE Computer Society (2004)
19. van der Aalst, W. M. P.: The Application of Petri Nets to Workflow Management. J. of Circuits, Systems and Computers. 8(1), 21--66 (1998)
20. van der Aalst, W. M. P., Basten, T., Verbeek, H.M.W., Verkoulen, P.A.C., Voorhoeve, M.: Adaptive Workflow: On the Interplay between Flexibility and Support. In: 1$^{st}$ Int. Conf. on Enterprise Information Systems, vol. 2, pp. 353–360. Kluwer Academic Publishers (1999)
21. Sadiq, S.W., Orlowska, M.E.: Analyzing process models using graph reduction techniques. Information Systems. 25(2), 117--134 (2000)
22. ter Hofstede, A.H.M., Orlowska, M.E., Rajapakse, J.: Verification Problems in Conceptual Workflow Specifications. Data Knowledge Engineering. 24(3), 239--256 (1998)
24. Dadam, P., Reichert, M.: The ADEPT Project: A Decade of Research and Development for Robust and Flexible Process Support. Tech. Report. Fakultät fur Ingenieurwissenschaften und Informatik, Ulm, http://dbis.eprints.uni-ulm.de/ 487/1/Reichert_01_09-2.pdf (2009)
25. Sadiq, S.W., Orlowska, M.E., Sadiq, W.: Specification and validation of process constraints for flexible workflows. Information Systems. 30(5), 349--378 (2005)